# Quasiphase transition of a single-file water chain influenced by atomic charges in a water model using orientational-biased replica exchange Monte Carlo simulations

Liang Zhao[a, †], Junqing Ni[a], Zhi Zhu[b], Yusong Tu[a] and Chunlei Wang[c,†]

[a]*College of Physical Science and Technology & Microelectronics Industry Research Institute, Yangzhou University, Jiangsu, 225009, China*
[b]*School of Optical-Electrical Computer Engineering, University of Shanghai for Science and Technology, Shanghai, 200093, China*
[c]*College of Sciences, Shanghai University, Shanghai, 200444, China*

†Corresponding authors: zhaoliang@yzu.edu.cn, wangchunlei1982@shu.edu.cn

**Abstract**
The recently observed temperature-dependent quasiphase transition of the single-file water chain confined within a carbon nanotube in experiments has been validated by the simple lattice theory and molecular dynamics simulations. It has been pointed out that the atomic charges in water models are important, yet how the values will affect the structural details and thermodynamic properties of the quasiphase transition has not been fully revealed. In this work, we perform orientational-biased replica exchange Monte Carlo simulations in the canonical ensemble to explore the effect of atomic charges in the SPC/E water model on the quasiphase transition of a single-file water chain. Based on the atomic charge values reported in literature, three distinct quasiphases are reproduced, comprising a fully hydrogen-bonded water chain at lower temperatures, a more ordered dipolar orientation along the tube axis at intermediate temperatures, and a completely disordered structure at higher temperatures. Then by increasing the atomic charge values, we find that the fragmentation of the entire water chain into shorter water segments, the orientational ordering of water dipoles along the tube axis, and the transition towards complete disorder are all inhibited. Consequently, the transition temperatures between three quasiphases have been shifted to higher temperatures. The thermodynamic analysis demonstrates that the increased atomic charge values enhance the hydrogen bonding between neighbouring water molecules also the electrostatic attraction within the water chain, leading to a longer water dipole correlation length even at higher temperatures. These findings highlight the vital role of atomic charges in water models also the electrostatic interaction in regulating the orientational ordering of water molecules under nanoconfinement.

## I. INTRODUCTION

Understanding the behavior of water confined within nanoscale environments [1-7], particularly in carbon nanotubes (CNTs) [8-11], provides valuable insights into the exploration of extraordinary transport properties [12-18] of nanopores or biological channels, as well as the manufacture of promising water-mediated applications such as signal conversion [19] and desalination devices [20]. Water molecules confined in CNT exhibit distinct structures and dynamic behaviors compared to those in bulk [10, 21]. When the CNT diameter decreases to less than 1 nm, the mutual passage of molecules is prohibited, and a single-file water chain is formed. The neighbouring water molecules are tightly connected by hydrogen bonds (HBs), and their dipoles are approximately aligned along the tube axis. Among various theoretical and computational investigations, the one-dimensional water chain model, despite its simplicity, offers insights into the interaction between water molecules and how these interactions influence the overall behavior of the water chain. Koflinger *et al.* introduced a one-dimensional (1D) dipole lattice model, in which each molecule is represented as a point dipole positioned at a regular lattice site [22-25]. The water chain can thus be viewed as several ordered segments consisting of equally oriented dipoles along the axis, separated by defects with vertical dipole orientations. The total energy is expressed as the sum of internal energies of the segments and their dipole-dipole type interactions. This model has been successful in reproducing the filling-empty transition, bistability of particle number distribution, and static dielectric response to an external field. A more realistic water chain model developed by Sahoo [26] and Serwatka *et al.* [27], treats water molecules as identical asymmetric rigid tops with fixed equal interdistances and orientational freedom in space, interacting through two-body or many-body water potentials [27]. Based on this model, quantum mechanical descriptions of the water chain, including HB ordering at shorter interdistances and dipolar ordering at larger distances, have been well reflected in the study of ground state behavior using the path integral ground state approach [26] or the density matrix renormalization group method [28]. When a more detailed description of the structure or dynamics of a single-file water chain is needed, such as the water gating and flipping [29-31], water pump [32], proton transfer process [12, 13, 33], the interaction sites of oxygen and hydrogen atoms should be explicitly considered and various water models have been employed in computational simulations.

Recently, a temperature-dependent quasiphase transition of the single-file water chain was first reported inside the (6, 5) single-walled carbon nanotube (SWCNT) by Ma *et al* [34]. They realized the direct comparison of photoluminescence spectra of empty and water-filled SWCNTs, and observed an additional stepwise spectrum shift for water-filled SWCNT at about 150 K. Using molecular dynamics (MD) simulations, they attributed the quasiphase transition to the temperature-dependent changes in orientational ordering of water dipoles. Druchok *et al.* further discussed in detail the existence of quasiphase transition by MD simulations of water molecules encapsulated inside (6, 5)-SWCNT [35]. They reproduced the quasiphase transition given by Ma *et al.*, most importantly, proposed a simple lattice model based on statistical mechanics. In this model, coplanar water dipoles were rendered on lattice sites of a straight line, with each site assigned three states subjected to configurational restrictions. The partition function of the water chain was obtained by the summation of contributions from the short-range nearest neighbour interaction and the long-range dipole-dipole interaction as well as the rotational entropy contributions. The quasiphase transition and the simple lattice model have been generalized to another polar molecule, ammonia, expanding the category of quasiphase transition [36]. In these works, the atomic charge in molecules is important and the values have been revised according to quantum chemistry calculations. For the water molecule, the calculated values are found to be significantly smaller than default values [34, 35], due to the coupling between the molecular orbitals of SWCNT and water molecules [37], as well as the lower coordination number in the one dimensional chain compared to that in bulk water [38]. For example, the atomic charge of oxygen atom $q_O$ in the extended simple point charge (SPC/E) water model is revised to $-0.4348e$, less than the default value of $-0.8476e$ [35]. However, the influence of atomic charges in water models on the quasiphase transition of a single-file water chain, such as the structural details and thermodynamic properties, has not been fully revealed.

To address this issue, we perform the orientational-biased replica exchange Monte Carlo (MC) simulation of a single-file water chain of the SPC/E water model with varied atomic charges. Based on the atomic charge values reported in literature, three distinct quasiphases are reproduced including a fully hydrogen-bonded water chain at lower temperatures, a more ordered dipolar orientation along the tube at intermediate temperatures, and a completely disordered structure at higher temperatures. Then, by varying the atomic charge values, we find that the increased atomic charge values inhibit the fragmentation of the entire water chain into shorter water segments, the orientational ordering of water dipoles along the tube axis, and the transition towards complete disorder. Consequently, the transition temperatures between three quasiphases have been shifted to higher temperatures. It is demonstrated that the larger atomic charge values enhance the hydrogen bonding between neighbouring water molecules also the electrostatic attraction within the water chain, leading to a longer dipole correlation length even at higher temperatures. These findings emphasize the importance of atomic charges of water models used in reproducing the quasiphase transition of the single-file water chain, offering a fundamental understanding of electrostatic interaction and hydrogen bonding in regulating the structures and dynamics of water molecules under nanoconfinement.

## II. MODEL AND METHODS

A chain of $N = 40$ water molecules is arranged along the one-dimension $x$-axis within a finite length $L = 24$ nm without periodic boundary condition, as shown in Fig. 1. Two hard spheres are fixed at the two endpoints to confine the water molecules within the interval. The movement of each water molecule is restricted along the $x$-axis for simplicity while the spatial orientation is free. Initially, water molecules are positioned and oriented randomly. The geometry of a water molecule is modeled by the rigid SPC/E model, where the H-O-H bond angle is 109°47' and the O-H bond length is 0.1 nm. The Lennard-Jones (LJ) parameters for oxygen are $\varepsilon_{OO} = 0.636$ KJ/mol and $\sigma_{OO} = 0.315$ nm, and oxygen-oxygen LJ interaction is described by $E_{LJ} = 4\varepsilon_{oo}\left[\left(\frac{\sigma_{oo}}{r}\right)^{12} - \left(\frac{\sigma_{oo}}{r}\right)^{6}\right]$. The LJ parameters of hydrogen atoms are neglected. The electrostatic interaction between two charges is given by $E_{ele} = \frac{1}{4\pi\varepsilon_0}\frac{q_i q_j}{r_{ij}}$, where $i, j$ denote the O or H atoms and the electrostatic charge $q_O = -Q$, $q_H = +0.5Q$. No cutoff radii are used in the calculations of LJ and electrostatic interactions. The value of $Q$ is varied from $0.2e$ to $0.8e$ with the increasement of $0.1e$, which covers a large range of charges obtained from different quantum calculations. The $Q = 0.4348e$ given by Druchok *et al*. [35] is taken as a reference system. The default value of $Q = 0.8476e$ is also considered for comparison.

The hydrogen bond (HB) is defined in terms of a geometric criterion: the O…O distance of a pair of water molecules is less than a cutoff distance $R_c$ and O-H…O angle is less than 30°. The $R_c$ is determined as the first minimum in the radial distribution function. There is a slight decrease of $R_c$ with the increase of $Q$, and we set $R_c = 0.40$ nm for $Q = 0.2e$, $0.3e$ and $0.4e$, $R_c = 0.35$ nm for $Q = 0.4348e$, $0.5e$, $0.6e$, $0.7e$, $0.8e$ and the default value of $0.8476e$ for accuracy (see radius distribution functions in PS. 1 of the Appendix). The HB energy of a pair of hydrogen-bonded water molecules is defined as the sum of LJ and electrostatic energies.

In the canonical ensemble (NVT) Monte Carlo simulations, three types of sampling algorithms are employed including traditional Metropolis-based translation and rotation, orientational-bias (OB) and replica exchange (RE) methods. OB is used to sample the configurations with HBs, which strongly depends on the relative molecular orientation [39]. RE shows the advantage in overcoming systems with possible multiple barriers [40] and outputting configurations at different temperatures in a single task submission. We have found that the combination of these algorithms is able to search the stable system states with lower energy and produce a more continuous sampling of potential energy, heat capacity and orientational ordering of the water chain (see comparisons of Metropolis sampling, Metropolis + OB and Metropolis + OB + RE for the reference system of $Q = 0.4348e$ in PS. 2 of the Appendix).

A single MC sweep consists of three types of MC moves: (i) Metropolis translational and orientational moves, (ii) Metropolis translational and orientational moves with the OB method, and (iii) RE swap. Trial moves (i) and (ii) are performed randomly within $2N = 80$ steps, followed by using one RE swap (iii). For the translational move, it is restricted along the *x*-axis and initial maximum allowed displacement of the oxygen atom is 0.015 nm. For the orientational move, one of three Cartesian axes is randomly chosen and initial maximum rotation angle is 5°. These two parameters are automatically adjusted to give the acceptance ratio of about 30%. For the OB method, 5 trial orientations are selected to construct the Rosenbluth factor. For the RE method, $M = 140$ replicas are used ranging from temperature $T = 1$ K to 279 K with the equal spacing of 2 K. Each replica exchanges the configuration with its two neighbours, that is, $M$ replicas can be paired up in two ways: [(1, 2), (3, 4), … ] or [(2, 3), (4, 5), … ] where the configurations of two replicas in the same parenthesis will be exchanged and the two ways are randomly chosen [41]. The total MC sweeps are $5\times10^6$ and the configurations in the last $1\times10^6$ sweeps are collected per 100 sweeps for analyses.

## III. RESULTS AND DISCUSSION
### A. Characterization of a single-file water chain

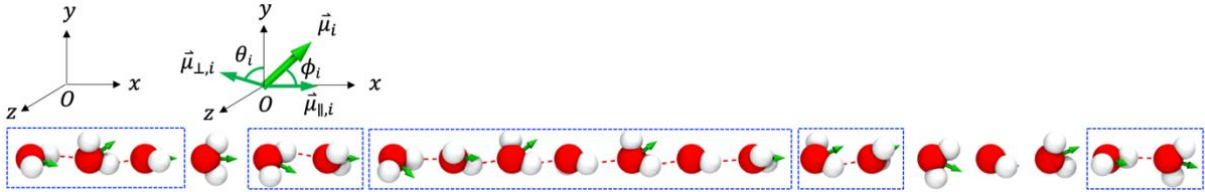

**FIG. 1.** Illustration of the configuration of a single-file water chain. Only twenty water molecules are shown here for display convenience. The green vector $\vec{\mu}_i$ represents the dipole of a water molecule indexed by $i$, with the tangent component $\vec{\mu}_{\parallel,i}$ (projection of $\vec{\mu}_i$ to the *x*-axis) and normal component $\vec{\mu}_{\perp,i}$ (projection of $\vec{\mu}_i$ into the *yOz* plane). Two polar angles $\phi_i$ and $\theta_i$ denote the angle between $\vec{\mu}_i$ and +*x* direction, and that between $\vec{\mu}_{\perp,i}$ and +*y* direction, respectively. The red dashed line connecting a hydrogen atom (white sphere) and an oxygen atom (red sphere) denotes the HB. The water molecules within a blue dashed frame represent a water segment, where neighbouring water molecules are connected by HBs.

In order to characterize the configuration of a single-file water chain, the following quantities are introduced: (I) A water segment is defined as a wire of water molecules where adjacent molecules are connected by HBs. According to this definition, water molecules are categorized into two groups: those within water segments and those not belonging to any water segment. Taking the configuration in Fig. 1 as an example, the whole water chain is divided into 5 water segments containing 16 water molecules in total and 4 individual water molecules out of any segment. The longest water segment is 7 water molecules' long; (II) The orientation of a water molecule indexed by $i$ is described by two polar angles, with $\phi_i$ denoting the angle between the dipole $\vec{\mu}_i$ and $+x$ direction and $\theta_i$ denoting the angle between $\vec{\mu}_{\perp,i}$ and $+y$ direction. The average dipole angle of water chain is given by $\bar{\phi} = \sum_{i=1}^{N} \phi_i /N$; (III) The normal component of dipole vector of individual water molecules is $\mu_{norm} = \overline{|\mu_\perp|}/\mu$, and the tangent component of the normalized total dipole vector of the water chain is $\mu_{tang} = \overline{\mu_\parallel^{total}}/(N\mu)$, where the overline represents the mean value. These two definitions are the same to those used in Ma's and Druchok's work; (IV) The average available space of a water molecule in water segments is defined by the average oxygen-oxygen distance of a HB, and that of a water molecule out of water segments is estimated by the half of the sum of oxygen-oxygen distances with its two neighbouring molecules.

## B. The quasiphase transition for $Q = 0.4348e$

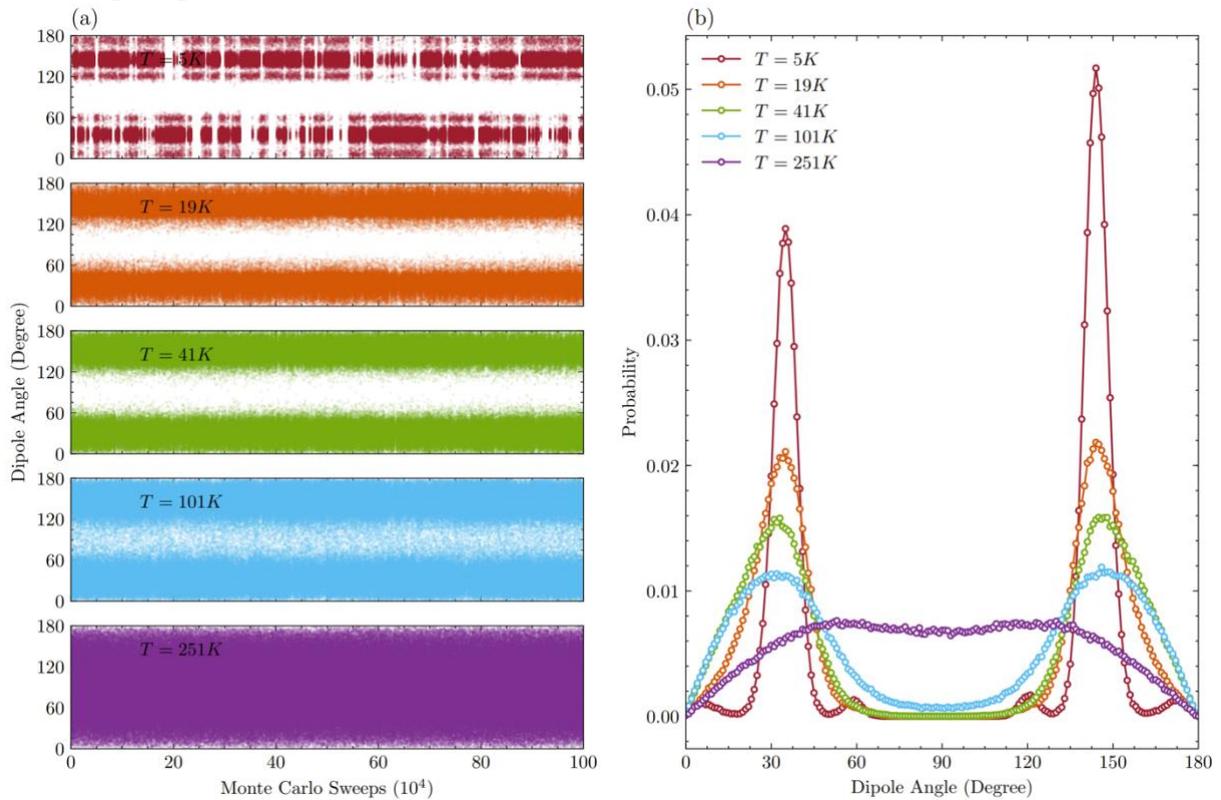

**FIG. 2.** (a) Scatter plots and (b) probability distributions of dipole angles of the water chain with respect to $+x$ direction at five typical temperatures $T$ = 5 K, 19 K, 41 K, 101 K and 251 K, for $Q = 0.4348e$. These temperatures correspond to the quasiphase I (5 K), quasiphase II (19 K, 41 K) and quasiphase III (101 K, 251 K), as will be discussed in Figs. 3 and 4.

It has been proved that the SPC/E water model with a revised $Q = 0.4348e$ shows the quasiphase transition in nanochannels, despite that this value is smaller than the default value of $0.8476e$. Therefore, we first check the phase behavior of the single-file water chain using $Q = 0.4348e$. Figure 2 shows the scatter plots and probability distributions of the water dipole angle $\phi_i$ at five typical temperatures. At a lower temperature $T$ = 5 K, the dipole angles mainly fall within two intervals of (0°, 60°) and (120°, 180°), representing water dipoles oriented closely along the $+x$ direction and its opposite direction. These symmetrical orientations are commonly achieved through water flipping. As the temperature increases, the boundary between these two intervals gradually diffuses and vanishes at a higher temperature $T = 251$ K, indicating the increased orientational freedom for water molecules. These changes become evident in the probability distribution of dipole angles shown in Fig. 2(b). Two sharp peaks are observed around 37° and 142° at $T = 5$ K. When $T$ increases, the peak positions remain unchanged while their widths broaden, resulting in a larger fluctuation in dipole orientation around both 37° and 142°. From ~ 101 K, the dipole orientation within the interval of (60°, 120°) emerges, and the dipole angle exhibits a relatively flat distribution at 251 K. This indicates that the ordered arrangement of water chain gradually diminishes from 101 K, an indication of the order-disorder transition.

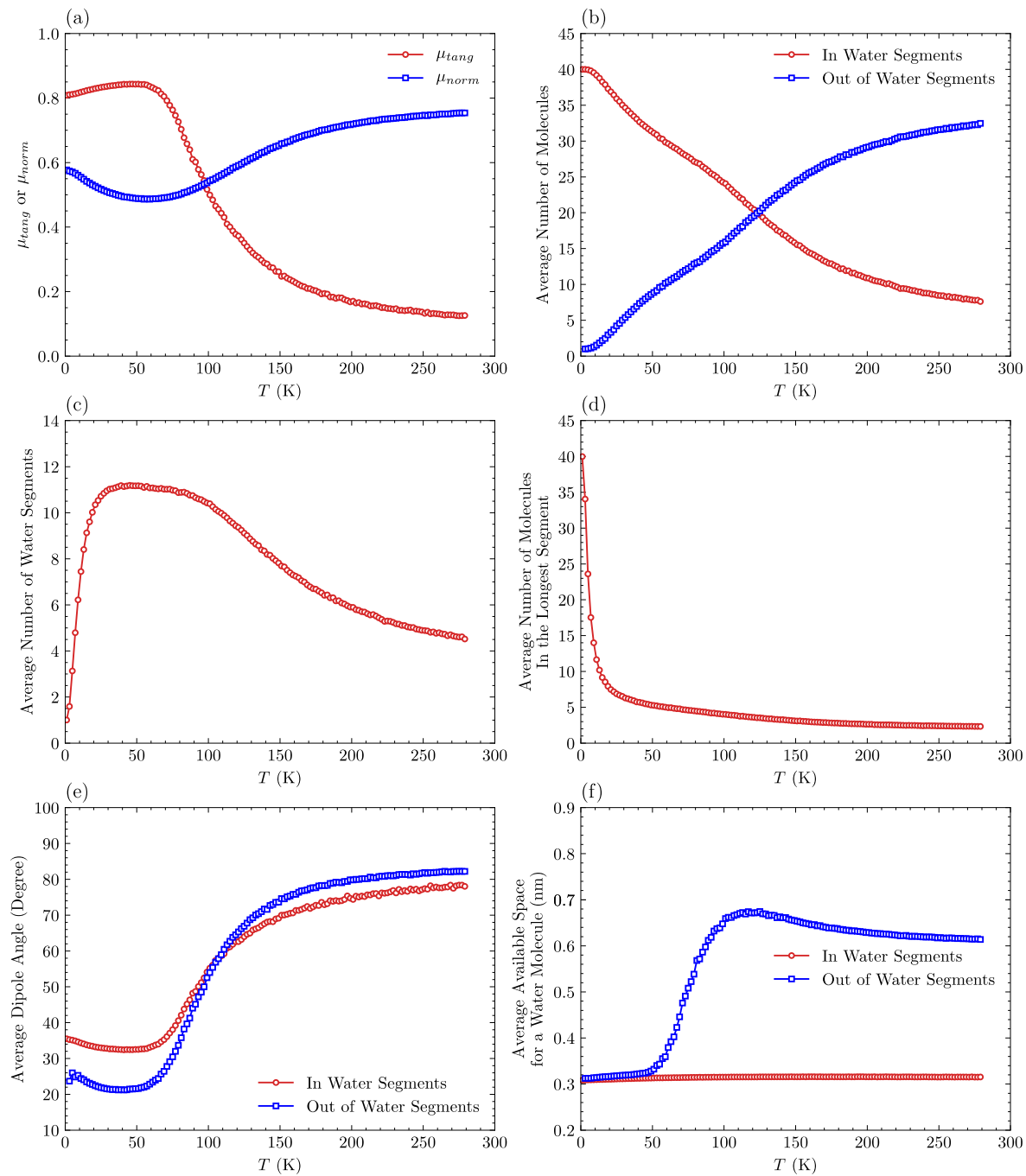

**FIG. 3.** (a) The average value of the normal component of individual water molecules $\mu_{norm}$ and the average value of tangent component of the total dipole of the water chain $\mu_{tang}$ for $Q$

= 0.4348$e$. Configurational details of the water chain with respect to temperature, including (b) average number of water molecules in and out of water segments, (c) average number of water segments, (d) average number of water molecules within the longest water segment, (e) average dipole angle of water molecules within and out of water segments, and (f) average available space for a water molecule in and out of water segments for $Q = 0.4348e$. These average values are averaged over water chain configurations under the condition $\bar{\phi} < 90°$ and the errors are shown by the shaded areas.

We first show the ordering of the whole water chain by calculating the $\mu_{tang}$ and $\mu_{norm}$. As discussed above, the water chain prefers two symmetric orientations, and we focus on the orientation along the +$x$ direction for convenience. The configuration is considered if the average dipole angle $\bar{\phi}$ is less than 90°. As shown in Fig. 3(a), $\mu_{tang}$ increases from $T = 1$ K and reaches the maximum at $T \sim 50$ K, accompanied by a decrease of $\mu_{norm}$ to the minimum. As $T$ continues increasing, the $\mu_{tang}$ experiences a rapid decay around 100 K. The trends observed in $\mu_{tang}$ and $\mu_{norm}$ are consistent with those given by Ma *et al.* (TIP3P water model with the revised charge) and Druchok *et al.* (SPC/E water model with the revised charge) via MD simulations.

Through a detailed analysis of the temperature-dependent water arrangement including dipole angle and water segment configuration, three distinct quasiphases denoted by I, II and III have been identified (A more reasonable estimation of temperature regimes for three quasiphases is given after the Fig. 4). In the quasiphase I at lower temperatures of several Kelvin, all water molecules are coordinated by HBs ($T = 1$ K, see Fig. 3(b)), forming a whole water segment (Fig. 3(c)). As the temperature increases closely to 10 K, $N \sim 39$ water molecules remain bonded by HBs (Fig. 3(b)). However, the whole water chain has been fragmented into 6 water segments (Fig. 3(c)), with the longest one containing an average of 14 water molecules (Fig. 3(d)). We note that the interdistance between water molecules is approximately 0.31 nm (Fig. 3(f)), which is less than the cutoff value for a HB. This means that the formation of multiple water segments is influenced by the reorientation of water molecules due to the increase of temperature. Indeed, the average dipole angle within water segments shows a slight decrease from 35.5° at 1 K to 34.3° at 11 K, and the dipole angle out of water segments also decreases to about 24.4° (Fig. 3(e)). This variation of water dipoles is consistent with the increase of $\mu_{tang}$ also the decrease of $\mu_{norm}$ in Figs. 3(a) and 3(b), indicating that water molecules are more oriented along the +$x$ direction. In the intermediate quasiphase II at around 50 K, the number of water molecules outside water segments increases to 9, and the number of water segments rises to 11, with a decrease in the number of water molecules in the longest segment to 5. This indicates that more water molecules escape from the coordination of HBs, leading to the formation of multiple shorter water segments. The whole water chain shows a more ordered structure in orientation, evidenced by the minima of 32.5° and 21.6° in Fig. 3(e) and also the extrema in $\mu_{tang}$ and $\mu_{norm}$ in Fig. 3(a). The interdistance between water molecules remains below the HB cutoff value. Upon the rise of temperature toward 100 K, we can see a notable increase of available space for water molecules out of water segments, from 0.33 nm to about 0.7 nm (Fig. 3(f)), accompanied by a dramatic increase of water dipole angle to about 60° (Fig. 3(e)). In the quasiphase III at higher temperatures above 100 K, nearly half of water molecules are out of water segments (Fig. 3(b)), with the number of water molecules in the longest segment decreasing to 2 (Fig. 3(d)). More water molecules are free of HBs and the disordered structure is formed. The significant loss of hydrogen bonds dominates the quasiphase transition. When $T$ reaches room temperature, water dipoles out of water segments tend to be oriented vertically to the $+x$ direction than those in water segments, with their values of 82° and 78°, respectively. Throughout the entire temperature range, the polar angle $\theta_i$ always has a uniform distribution (see PS. 3 in Appendix) due to the symmetry of the system.

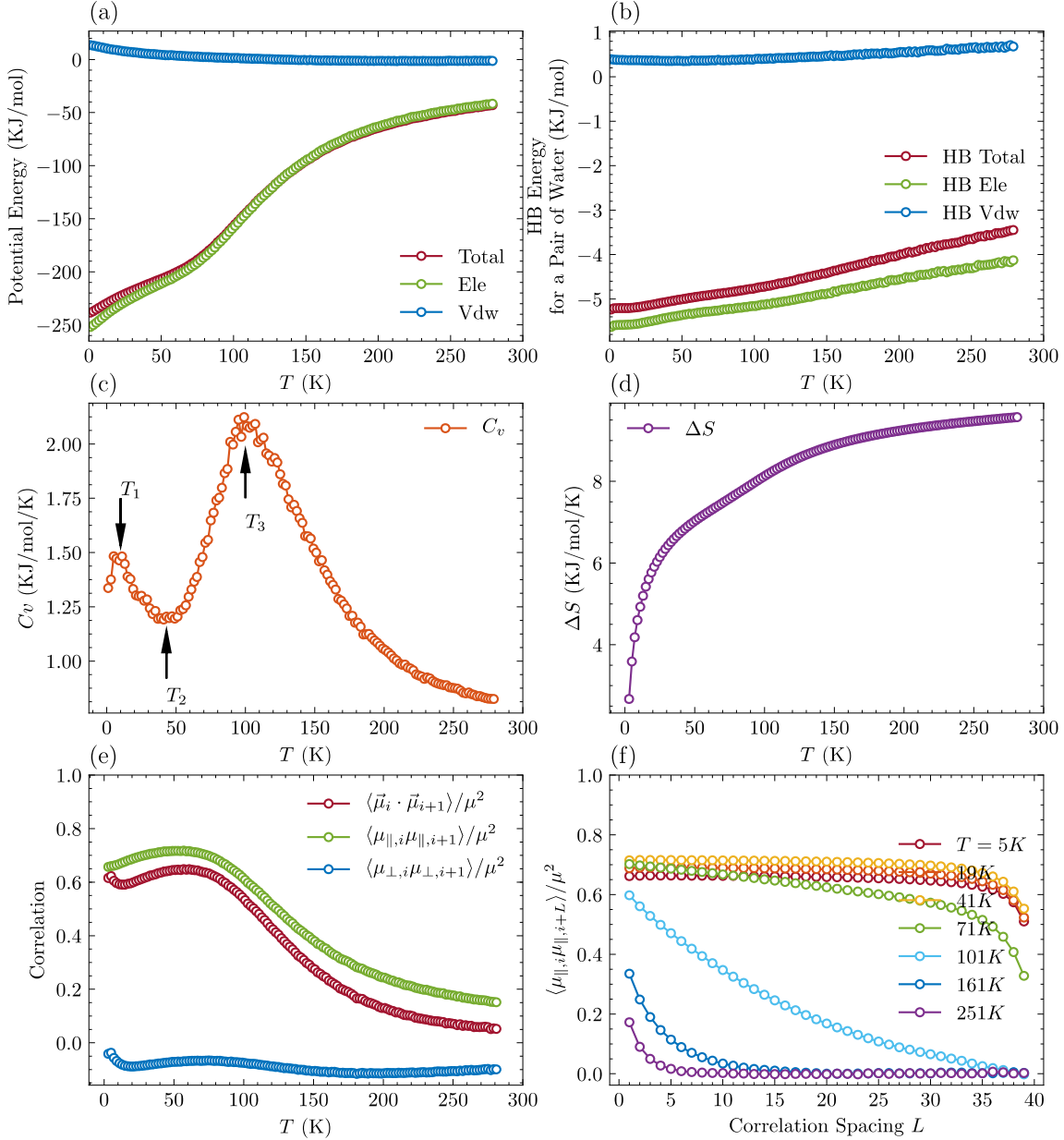

**FIG. 4.** Temperature-dependent thermodynamic quantities and correlations for $Q = 0.4348e$. (a) Average total potential energy ($E_p$) of water chain decomposed into contributions from electrostatic ($E_{ele}$) and van der Waals energies ($E_{vdw}$) under $E_p = E_{ele} + E_{vdw}$. (b) Average HB energy between a pair of water molecules decomposed into contributions from electrostatic and van der Waals energies. (c) Heat capacity ($C_v$) and (d) an estimated entropy ($\Delta S$) relative to the entropy at $T = 1$ K of the single-file water chain. (e) Correlations between neighbouring water molecules, including the correlations of dipole vector $\vec{\mu}_i$, tangent component $\mu_{\parallel,i}$ and normal component $\mu_{\perp,i}$. (f) Temperature dependence of correlation of $\mu_{\parallel,i}$ on two water molecules with the spacing $L$ denoting the difference between indices of two molecules. Black arrows with $T_1$, $T_2$ and $T_3$ in Fig. 4(c) are used to denote the temperature regimes of three quasiphases.

With the description of structure and transition of quasiphases of the single-file water chain, now we turn to the temperature-dependent thermodynamic quantities and correlations. As shown in Fig. 4(a), the total potential energy shows an increase from about -250 KJ/mol to -50 KJ/mol across the temperature range primarily driven by electrostatic interactions, while the van der Waals interaction shows a weak repulsion with negligible positive values. Similarly, the HB energy between water pairs, increasing from -5.2 KJ/mol to -3.4 KJ/mol, is predominantly influenced by electrostatic attraction (Fig. 4(b)).

We further calculate the heat capacity by the expression $C_V = C_V^k + C_V^p = \frac{1}{2}RfN + \frac{\langle E_p \rangle^2 - \langle E_p^2 \rangle}{RT^2}$ [42], where the first term with freedom $f = 4$ corresponds to the contribution from kinetic energy related to one translational freedom along the $x$-axis and three spatial rotational freedoms, and the second term is from the fluctuation of total potential energy. Fig. 4(c) shows two peaks at $T \sim 10$ K and 100 K, dividing the temperature range into three regimes corresponding to three quasiphases. The presence of the first peak at lower temperatures is noteworthy, as the heat capacity usually a constant (in the ideal gas model) or increases with the increase of temperature (in the Einstein or Debye solid model). According to the structural analysis in Fig. 3, the whole HB bonded water chain breaks into a few water segments at $T \sim 10$ K and forms a more ordered orientation along the $+x$ axis meantime keeping the interdistance between water molecules unchanged. We therefore assume that the peak is Schottky-like heat capacity which is likely due to the excitation of rotational freedom in a few water molecules. The second peak is at $T \sim 100$ K, signifying the formation of disordered structure, as indicated in Fig. 3. In fact, the relative entropy $\Delta S$ with reference to $T = 1$ K, calculated by $\Delta S(T) = \int_1^T \frac{C_V}{T} dT$, continuously increases across the whole temperature range (Fig. 4(d)). This suggests that the structure of the single-file water chain continuously evolves toward disorder despite the existence of an ordered dipolar orientation of water molecules in quasiphase II. In terms of these analyses of $C_v$ profile and the $\mu_{tang}$ in Fig. 3(a), temperature regimes of three quasiphases can be more reasonably defined: quasiphase I at $T < T_1 \sim 10$ K, where $T_1$ is the position of the first peak in $C_v$ profile. In quasiphase I, almost all the water molecules are hydrogen bonded; quasiphase II in the vicinity of $T_2 \sim 50$ K, where $T_2$ denotes the minimum in $C_v$ or the maximum in $\mu_{tang}$ profile. In this regime, water dipoles are preferably orientated along the tube axis; quasiphase III at $T > T_3 \sim 100$ K, where $T_3$ refers to the second peak in $C_v$ profile. In this quasiphase, the disordered state of the water chain appears.

Figure 4(e) shows the correlations between dipoles of two water molecules. The positive correlation of $\mu_{\parallel,i}$ and the negative correlation of $\mu_{\perp,i}$ indicate the staggered arrangement of neighbouring water dipoles. For $T < 10$ K, the decrease of $\mu_\perp$ correlation dominates the $\vec{\mu}$ correlation despite that there is an increase in $\mu_\parallel$ correlation. This confirms that neighbouring water dipoles are more aligned along the $x$-axis, as shown in Fig. 3(e). Meantime, there is an increased relative rotation on $x$-axis, making a more negative value of $\mu_\perp$ correlation (see the relative angle of $\mu_\perp$ of two neighbouring water dipoles in PS. 4 of the Appendix). When $T$ is in the vicinity of 50 K, the water chain reaches the most ordered dipolar orientation, as evidenced by a peak in both $\mu_\parallel$ correlation and $\vec{\mu}$ correlation. For $T \sim 100$ K, we can see the rapid decrease in $\mu_\parallel$ and $\vec{\mu}$ correlations, indicating the loss of HBs and a more disordered structure. The correlation of $\mu_\parallel$ between neighbouring water molecules predominantly influences the water chain across the whole temperature range. As shown in Fig. 4(f), the $\mu_\parallel$ correlation can nearly extend over the whole water chain at temperatures of 5 K (quasiphase I), 19 K (quasiphase II), 41 K (quasiphase II), 71 K (quasiphase II). A substantial decrease can be found at $T = 101$ K (quasiphase III), and when $T = 251$ K (quasiphase III) the correlation spacing is reduced to about only 5 water molecules.

## C. The quasiphase transition for different values of $Q$

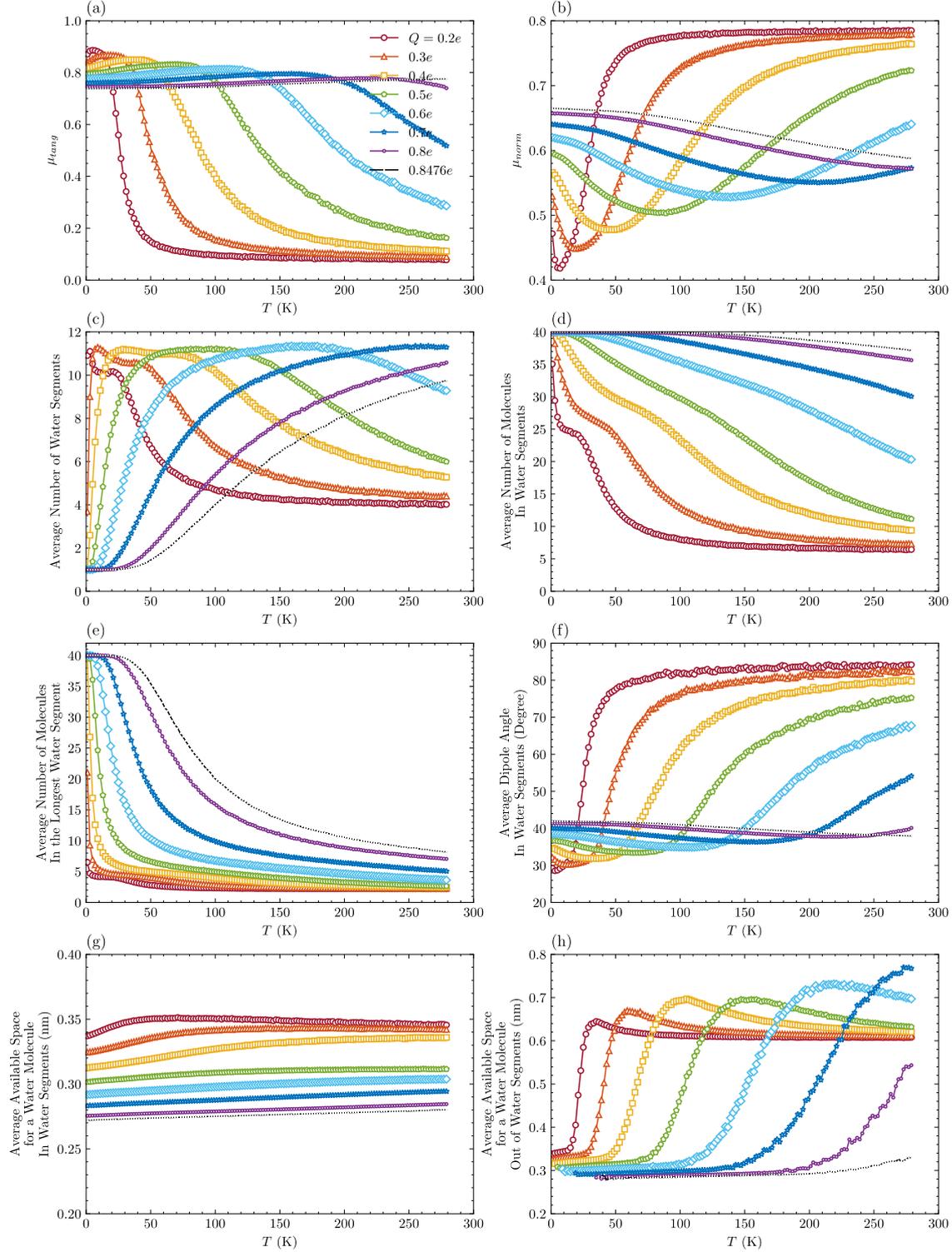

**FIG. 5.** Structural changes of the single-file water chain with respect to temperature, for $Q$

ranging from 0.2$e$ to 0.8$e$ with the increasement of 0.1$e$ and the default value of $Q$ = 0.8476$e$. (a) $\mu_{tang}$, (b) $\mu_{norm}$, (c) average number of water segments, (d) average number of water molecules in water segments, (e) average number of water molecules in the longest water segment, (f) average dipole angle in water segments, average available space for a water molecule (g) in water segments and (h) out of water segments. In (h), a few dots at lower temperatures are missing as all the water molecules are coordinated by HB and no water molecules are out of water segments.

We now turn to the quasiphase transition and structural changes under different atomic charges of oxygen atom in water model. As shown in Figs. 5(a) and 5(b), for $Q$ within the range from 0.2$e$ to 0.8$e$, $\mu_{tang}$ ($\mu_{norm}$) all show the extrema. For $\mu_{tang}$, the maxima are at around 5 K (0.2$e$), 17 K (0.3$e$), 43 K (0.4$e$), 71 K (0.5$e$), 119 K (0.6$e$), 159 K (0.7$e$) and 223 K (0.8$e$). This indicates that the orientational ordering of water dipoles in quasiphase II can exist in a wide range of atomic charge values. For larger $Q$ values, the water dipoles in quasiphase II exhibit a more staggered arrangement, resulting in reduced maxima values in $\mu_{tang}$ (increased minima values in $\mu_{norm}$) also the larger values of minima in the average water dipole angle for water molecules in water segments (Fig. 5(f)). For smaller $Q$ values, we can see the rapid decrease in $\mu_{tang}$ at lower temperatures, an indication of a narrower temperature range for the loss of ordering towards complete disorder around quasiphase III. The increase of $Q$ evidently inhibits the rapid fragmentation of the single-file water chain into water segments (Fig. 5(c)) also the rapid decrease of the average number of water molecules in the longest water segment (Fig. 5(d)). For larger $Q$ values, such as 0.8$e$, all water molecules remain in a whole water chain even when $T \sim 40$ K (Figs. 5(c), 5(d) and 5(e)). The larger $Q$ values also lead to a much shorter interdistance between neighbouring water molecules in water segments, for example, 0.27 nm for 0.8$e$, less than 0.35 nm for smaller $Q$ = 0.2$e$ (Fig. 5(g)). The peaks in available space for water molecules out of water segments in Fig. 5(h) has been shifted towards higher temperatures. It is noteworthy that for $Q$ = 0.8$e$, the difference of $\mu_{tang}$ between its maximum and the value at $T$ = 1 K becomes negligible, an implication of the disappearance of quasiphase II. For a larger value, i.e., the default value of $Q$ = 0.8476$e$, we cannot see the maximum in $\mu_{tang}$ over the current temperature range. It has been reported that the water chain evolves from the hydrogen-bonded arrangement at lower temperatures directly to the disordered structure for the default value [35].

The structural variations of the single-file water chain with respect to temperature at different values of $Q$ suggests the significant influence of electrostatic interactions in quasiphase transitions. At a fixed temperature, the larger values of $Q$ result in the lower total potential energy (dashed lines in Fig. 6(a)) and HB energy (Fig. 6(b)). For $Q \geq 0.5e$, the total potential energy is higher than the electrostatic energy within the whole temperature range, indicating that the electrostatic attraction dominates the interaction among water molecules such that the van der Waals interaction becomes repulsive. The increased electrostatic interaction pushes the quasiphases II, III and order-disorder transition to higher temperatures, as evidenced by the right shift of the minima in Fig. 6(c) from 9 K (0.2$e$) to 181 K (0.8$e$) and peaks from 30 K (0.2$e$) to 200 K (0.6$e$). The peaks for $Q$ = 0.7$e$, 0.8$e$ and 0.8476$e$ are estimated to be above 279 K. In addition, the Schottky-like peak at lower temperatures, which is assumed to be induced by the excitation of rotational freedom from a completely HB bonded water chain, shifts from several Kelvin for smaller $Q$ to about 30 K for $Q$ = 0.8$e$. Despite the more ordered dipolar orientations in quasiphase II, the entropy of the single-file water chain in fact monotonically increases with respect to temperature for all $Q$ values (Fig. 6(d)). The increased electrostatic attraction also moves the peak positions of correlation of tangent component $\mu_\parallel$ of two neighbouring water molecules (Fig. 6(e), see PS. 5 in the Appendix for correlations of dipole vector $\vec{\mu}_i$ and of normal component $\mu_\perp$) to higher temperatures, and significantly extends the correlation from several water molecules to the whole water chain (Fig. 6(f), see PS. 5 in the Appendix for other fixed temperatures).

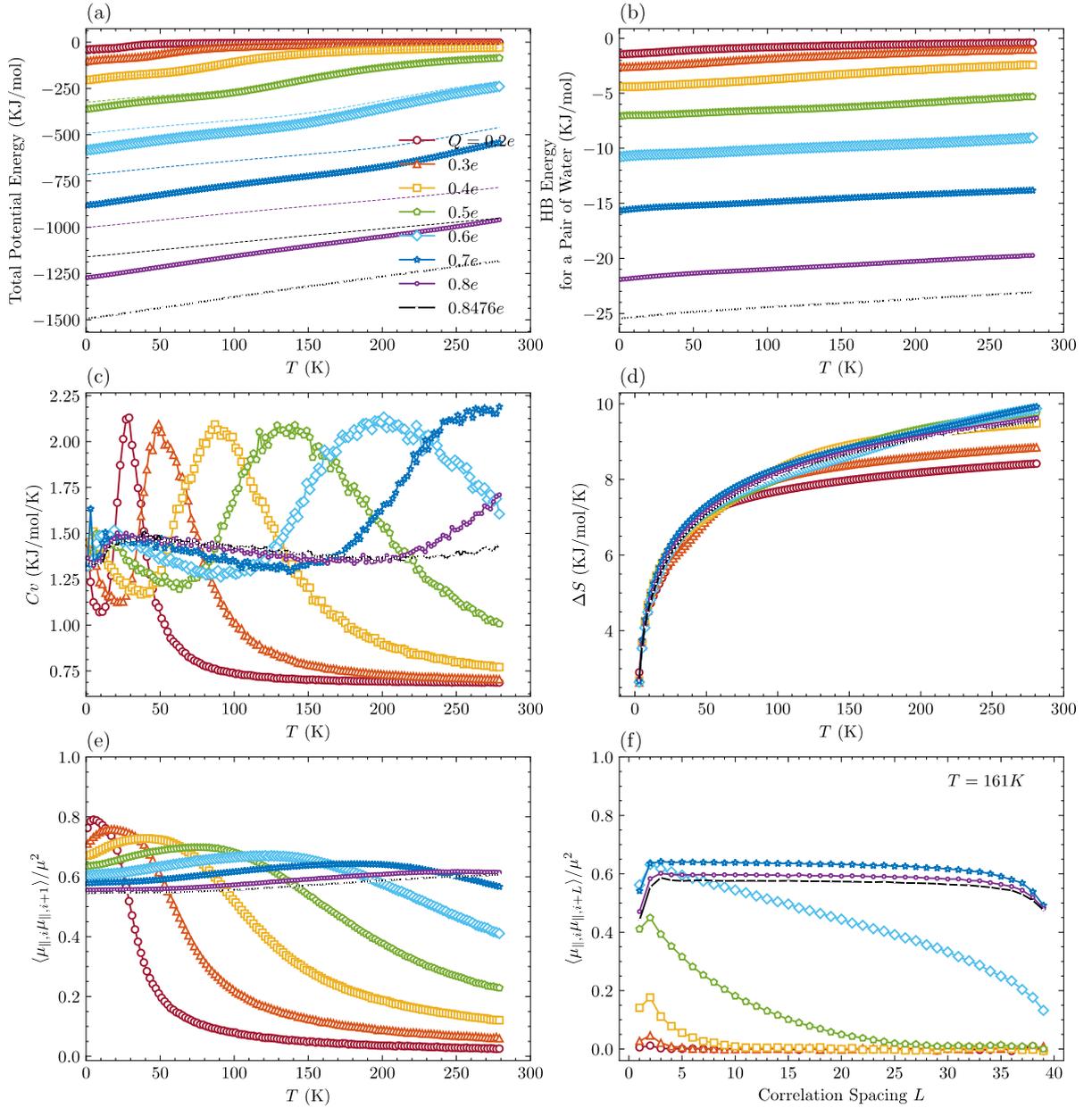

**FIG. 6.** Thermodynamic quantities of the single-file water chain with respect to temperature, for $Q$ ranging from $0.2e$ to $0.8e$ with the increasement of $0.1e$ and the default value of $Q = 0.8476e$, including (a) total potential energy (dashed lines) and electrostatic energy (dotted lines), (b) HB energy for a pair of water molecules, (c) heat capacity $C_v$, (d) entropy changes $\Delta S$ with respect to $T = 1$ K, (e) correlation of tangent component $\mu_\parallel$ and (f) dependence of correlation of $\mu_\parallel$ on two water molecules with the spacing $L$ at a fixed temperature of 161 K.

### D. Comparison with experimental data

The close dependence of transition temperatures on the atomic charges manifests that the atomic charges used in the SPC/E water model in classical simulations should be carefully considered with reference to experimental data. In fact, many experimental techniques, such as infrared spectrum [43], Raman spectroscopy [44], nuclear magnetic resonance [45-47], X-ray diffraction [48], neutron scattering [49] and dielectric relaxation [50], have been applied to the investigation of structural phase transition or dynamics of water confined within SWCNT with different chiralities and diameters. However, the water filling of SWCNT in these experiments is not in a single-file manner due to the large diameter greater than 1 nm. With the realization of the single-file water filling [34, 51], we use our results as a benchmark to estimate which atom charge is more reasonable.

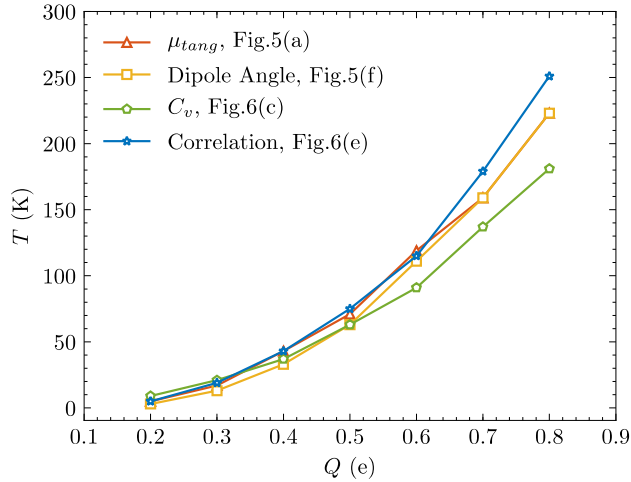

**FIG. 7.** The temperature of quasiphase II with respect to the atomic charge of oxygen atom in SPC/E water model. The value of temperature is determined from four types of quantities including $\mu_{tang}$ (Fig. 5(a)), the average dipole angle in water segments (Fig. 5(f)), the heat capacity (Fig. 6(c)) and the correlation of tangent component of neighboring water molecules (Fig. 6(e)).

Figure 7 shows the temperature of quasiphase II with respect to the value of $Q$, which ranges from $0.2e$ to $0.8e$. It can be seen that the temperatures of quasiphase II obtained from four quantities are different. This is attributed to the fact that the structural change of the single-file water chain is not a real phase transition but a continuous structural transformation between quasiphases. Thus, the exact determination of the temperature is difficult. According to the experimental data that an additional shift in PL spectrum of the single-file water filled SWCNT is at ~ 150 K [34], the value of $Q$ in the SPC/E water model is estimated to be around $0.7e$, which is smaller than the default value of $0.8476e$. Nonetheless, we still note that the pressure effect has not been taken in account in our simplified water chain model, which may also have impact on the estimation of the value of $Q$ in comparison with experimental data. A detailed discussion will be addressed in our forthcoming analyses.

### IV. CONCLUSIONS

In summary, we have performed orientational-biased replica exchange Monte Carlo simulations in the canonical ensemble to explore the quasiphase transition of a single-file water chain affected by the atomic charges in the SPC/E water model. It has been demonstrated that the atomic charges play the critical role in the quasiphase transition of the single-file water chain. Using the atomic charge as reported in literature, *i.e.*, $q_O$ = -0.4348$e$, we have reproduced three distinct quasiphases comprising a fully hydrogen-bonded water chain at lower temperatures, a more ordered dipolar orientation along the tube at intermediate temperatures, and a completely disordered structure at higher temperatures. Then, we vary the atomic charge of $q_O$ from -0.2$e$ to -0.8$e$ to explore how these values will affect the structural details and thermodynamic properties of quasiphase transitions. With the increase of magnitude of $q_O$, we find that the fragmentation of the entire water chain into shorter water segments, the orientational ordering of water dipoles, and the transition towards complete disorder are all inhibited. Consequently, the transition temperatures between three quasiphases have been shifted to higher temperatures. The thermodynamic analysis demonstrates that the larger atomic charges enhance the hydrogen bonding between neighbouring water molecules also the electrostatic attraction within the water chain, leading to a longer water dipole correlation length even at higher temperatures. Our simplified water chain model, where water molecules are restrained in one dimension, provides the fundamental understanding of the critical role of atomic charge values as well as the electrostatic interaction, which would be beneficial for the study of structures and dynamics of polar molecules under nanoconfinement.

## ACKNOWLEDGEMENTS

We acknowledge the valuable discussion on the programming with Dr. Nan Sheng. This work was supported by the National Natural Science Foundation of China [Nos: 11605151, 12074394, 12075201]; the National Science Fund for Outstanding Young Scholars (No. 12022508). L. Zhao also acknowledge the computational resources provided by the Hefei Advanced Computing Center.

# APPENDIX

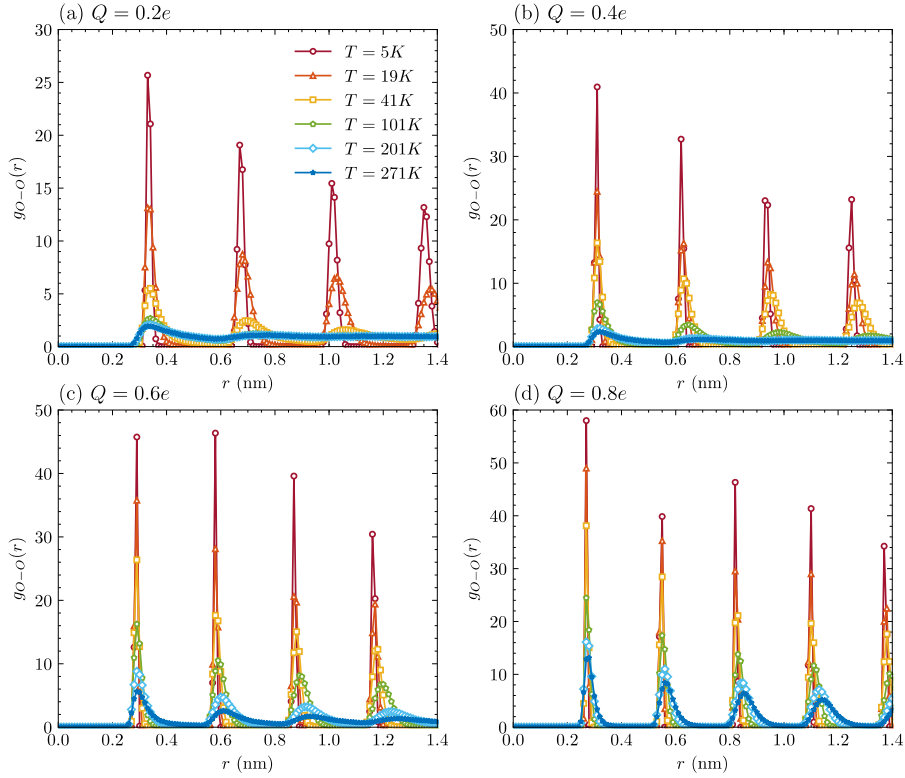

**PS. 1.** Radius distribution functions of two oxygen atoms, $g_{O-O}(r)$, at various temperatures under different $Q$ values. The $g_{O-O}(r)$ is defined by $g_{O-O}(r) = \frac{1}{\bar{\rho}} \frac{\langle n(r) \rangle}{\Delta r}$, where $\langle n(r) \rangle$ is the ensemble average number of oxygen atoms in the interval $[r, r + \Delta r]$ and $\bar{\rho} = \frac{N}{L}$ is the average number density. The increment $\Delta r$ is 0.01 nm. To identify the short-range arrangement of water molecules, we only show the $g_{O-O}(r)$ in the distance range less than 1.4 nm.

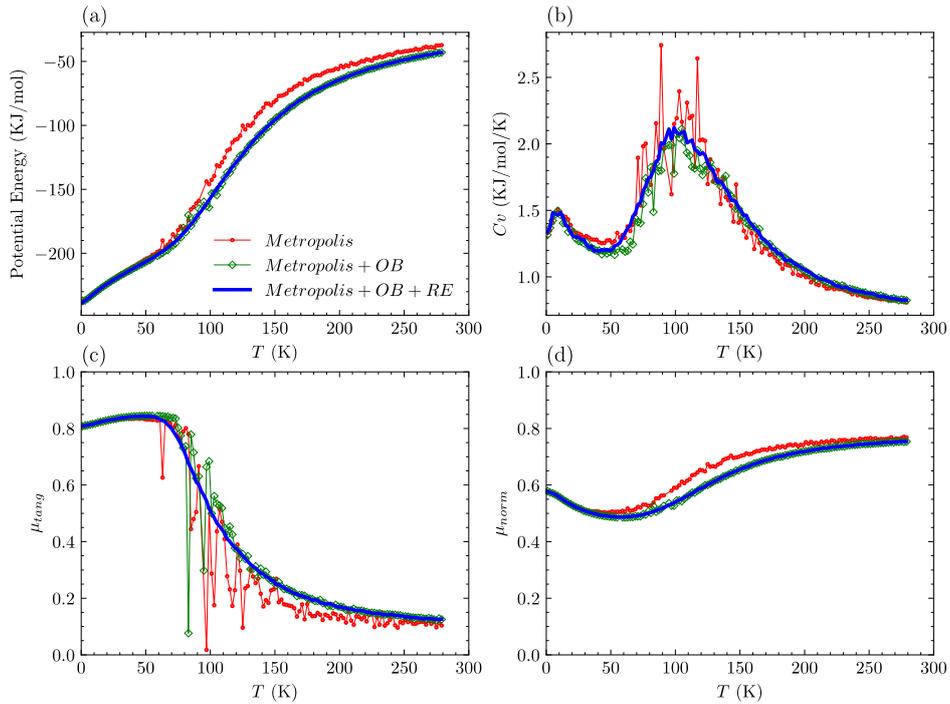

**PS. 2.** Comparisons of analysis of (a) potential energy, (b) heat capacity $C_v$, (c) $\mu_{tang}$ and (d) $\mu_{norm}$ for $Q = 0.4348e$ using three different Monte Carlo sampling methods including

Metropolis-based translation/rotation algorithm (Metropolis), Metropolis translation/rotation plus the orientational bias methods (Metropolis + OB), and Metropolis translation/rotation plus the orientational bias and replica exchange methods (Metropolis + OB + RE). For Metropolis method, data are missing at $T$ = 93 K and 95 K due to that the average dipole angle of the single-file water chain $\bar{\phi} > 90°$. Similar cases are at $T$ = 57 K and 93 K for Metropolis + OB method.

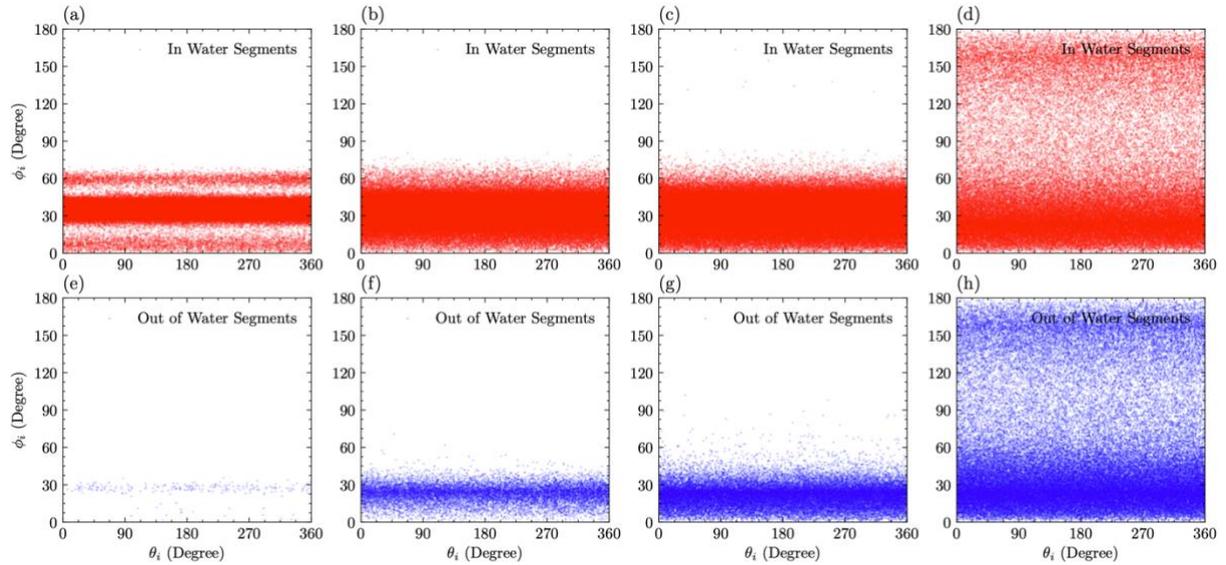

**PS. 3.** Orientational maps of two polar angles, $\theta_i$ and $\phi_i$, for water molecules with $Q = 0.4348e$ at (a, e) $T$ = 5 K, quasiphase I, (b, f) 19 K, quasiphase II, (c, g) 41 K, quasiphase II and (d, h) 101 K, quasiphase III. Upper panel: in water segments; Bottom panel: out of water segments.

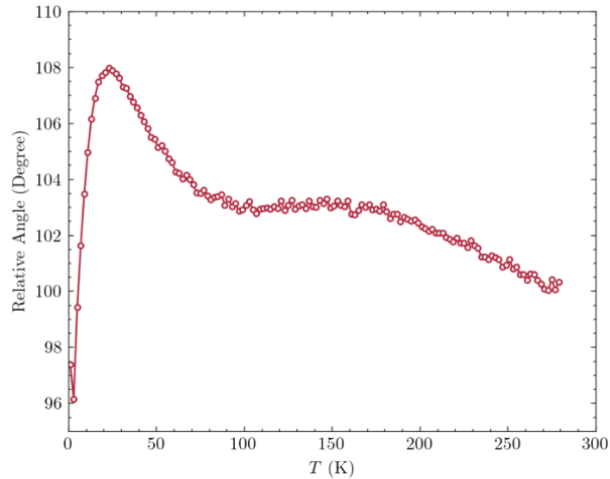

**PS. 4.** Relative angle of normal components $\mu_\perp$ of two neighbouring water dipoles for $Q$ = 0.4348$e$ at different temperatures.

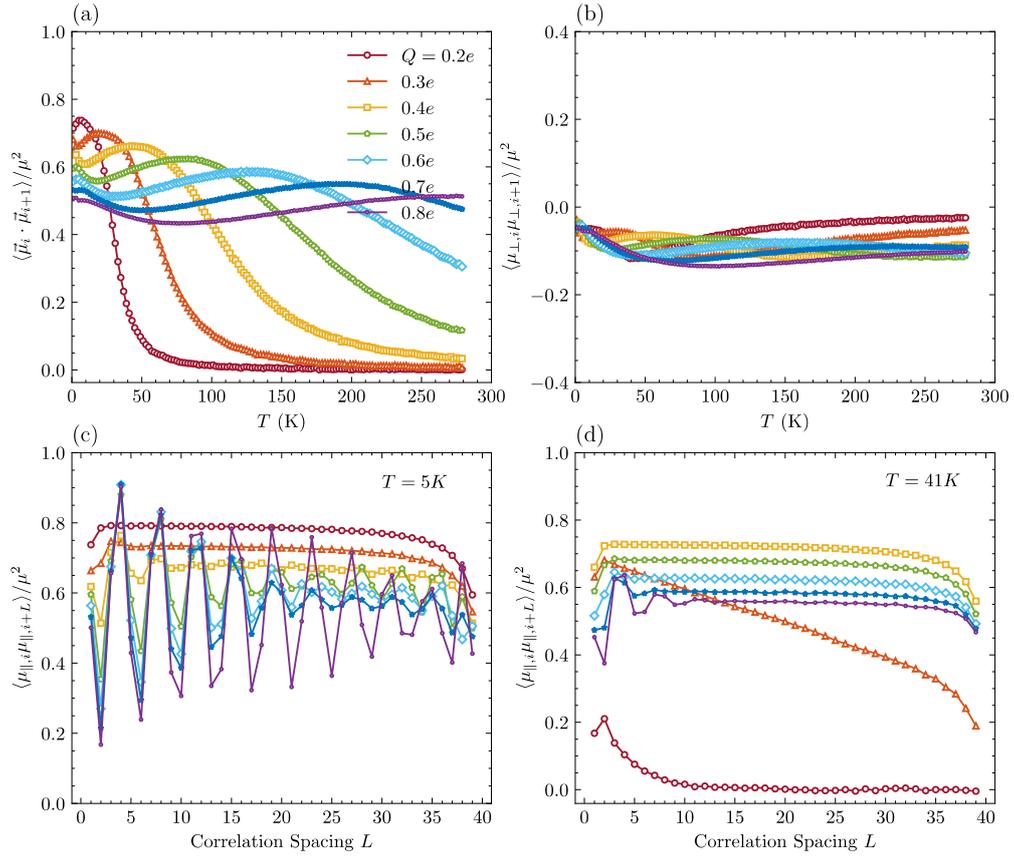

**PS. 5.** Correlations between neighbouring water molecules, including the correlations of (a) dipole vector $\vec{\mu}_i$ and of (b) normal component $\mu_\perp$. (c) Dependence of correlation of $\mu_\parallel$ on two water molecules with the spacing $L$ at fixed temperatures of 5 K and 41 K. $L$ represents the difference between indices of two molecules.